\documentclass[aps,tightenlines,twocolumn,floats,prd,superscriptaddress,floatfix]{revtex4-1}
\usepackage{graphicx}
\usepackage{epsf} 
\usepackage{amsfonts}
\usepackage{amsmath}
\usepackage{amssymb}
\usepackage{bbold}
\usepackage{bm}
\usepackage{bbm}
\usepackage{slashed}
\usepackage{dcolumn}

\begin{document} 

\title{Linear confinement in momentum space: singularity-free bound-state equations}

\author{Sofia Leit\~ao}
\email{sophia.leitao@gmail.com}
\affiliation{Departamento de F\'isica, Universidade de \'Evora, 7000-671 \'Evora, Portugal}
\affiliation{Centro de F\' isica Te\' orica de Part\' iculas (CFTP), Instituto Superior T\'ecnico (IST), Universidade de Lisboa, 
1049-001 Lisboa, Portugal}

\author{Alfred Stadler}
\email{stadler@uevora.pt}
\affiliation{Departamento de F\'isica, Universidade de \'Evora, 7000-671 \'Evora, Portugal}
\affiliation{Centro de F\' isica Te\' orica de Part\' iculas (CFTP), Instituto Superior T\'ecnico (IST), Universidade de Lisboa, 
1049-001 Lisboa, Portugal}

 \author{M. T. Pe\~na}
\email{teresa.pena@tecnico.ulisboa.pt}
\affiliation{Centro de F\' isica Te\' orica de Part\' iculas (CFTP), Instituto Superior T\'ecnico (IST), Universidade de Lisboa, 
1049-001 Lisboa, Portugal}
\affiliation{Departamento de F\'isica, Instituto Superior T\'ecnico (IST), Universidade de Lisboa, 
1049-001 Lisboa, Portugal}
 
 \author{Elmar P. Biernat}
 \email{elmar.biernat@tecnico.ulisboa.pt}
\affiliation{Centro de F\' isica Te\' orica de Part\' iculas (CFTP), Instituto Superior T\'ecnico (IST), Universidade de Lisboa, 
1049-001 Lisboa, Portugal}

\date{\today}

\begin{abstract}
Relativistic equations of Bethe-Salpeter type for hadron structure are most conveniently formulated in momentum space. The presence of confining interactions causes complications because the corresponding kernels are singular. This occurs not only in the relativistic case but also in the nonrelativistic Schr\"odinger equation where this problem can be studied more easily. For the linear confining interaction  the singularity  reduces to one of Cauchy principal value form. Although this singularity is integrable, it still makes accurate numerical solutions difficult. We show that this principal value singularity can be eliminated by means of a subtraction method. The resulting equation is much easier to solve and yields accurate and stable solutions.  To test the method's numerical efficiency, we performed a three-parameter least-squares fit of a simple linear-plus-Coulomb potential to the bottomonium spectrum.
\end{abstract}

\maketitle

\section{Introduction}

In the simple but very successful older nonrelativistic quark models, mesons are described as bound states of constituent quark-antiquark pairs, interacting through a long-range confining linear potential and a short-range color-Coulomb potential. These models, often variations of the Cornell potential models \cite{Eichten:1975,Eichten:1978,Eichten:1980}, are able to explain a large variety of phenomena in meson spectra and decay rates. However, they also have a number of shortcomings, among which we highlight that they ignore the dynamical structure of constituent quarks, whose self-interaction gives rise to a momentum-dependent mass, as well as its connection to dynamical chiral symmetry breaking. 

Another weakness is that relativity is either not taken into account at all or only in a rudimentary fashion. An example is the well-known ``relativized'' quark model by Godfrey and Isgur \cite{Godfrey:1985aa} that only includes corrections from relativistic kinematics. 
A nonrelativistic framework is probably sufficient to explain most phenomena in heavy quarkonia, but systems with at least one light quark certainly require a relativistic treatment. 

In two recent papers \cite{Biernat:2014a,Biernat:2014b} we initiated a program on the theory of meson structure that continues and improves on previous work by Gross and Milana \cite{Gross:1991te,Gross:1991pk,Gross:1994he} and Savkli and Gross \cite{Savkli:1999me} using the Covariant Spectator Theory (CST) \cite{Gro69,Gro82,Gro82b,Sta11}. The goal of this program is the construction of a relativistically covariant model for all mesons that can be understood as quark-antiquark pairs, and which is made self-consistent by calculating the quark self-energy from the same interaction kernel that describes the quark-antiquark pair interaction.

Nonperturbative covariant descriptions of meson structure are also provided by the approaches based on the Dyson-Schwinger formalism \cite{Qin2011,Qin2012,Maris:2003vk,Alkofer:2000wg,Fischer:2006ub}. However, whereas those are usually formulated in Euclidean space, the equations of the CST are solved in Minkowski space, which enables us to calculate form factors without having to worry about uncertainties in extrapolations from the unphysical to the physical region.

In the CST model of Refs.~\cite{Biernat:2014a,Biernat:2014b}, the confining quark-antiquark interaction kernel is taken as a covariant generalization of the nonrelativistic linear potential, to which a constant is added (this is for the sake of simplicity -- at a later stage it will be replaced with a one-gluon exchange interaction). 

Fully relativistic equations require a momentum-space formulation. This makes it easier than in coordinate space to include a running coupling constant in the one-gluon exchange interaction, but more difficult to deal with the linear interaction which takes on the highly singular form of a double pole. It turns out that by reorganizing the momentum-space equations this singularity can be weakened to one of Cauchy principal value type, which is integrable but still quite cumbersome in practical applications. In order to construct a model that fits the whole meson spectrum, we need to have an accurate and stable numerical method to solve the CST equation with such a singular kernel at our disposal. Moreover, it has to be fast enough to make a least-$\chi^2$ fit feasible.

In this paper, we address the problem of solving the momentum-space CST equations with the linear interaction in its nonrelativistic limit, in which the CST equation becomes the Schr\"odinger equation. This is useful for several reasons: (i) The type of singularity in the nonrelativistic potential is the same as in the relativistic kernel. An efficient method to solve the Schr\"odinger equation can then be applied to the CST equation as well. (ii) For $S$-waves, the analytic solution of the Schr\"odinger equation with a linear potential is known. The energy eigenstate wave functions in coordinate space are given in terms of Airy functions, and the energy eigenvalues through their roots. This represents an ideal test case for evaluating numerical methods. (iii) Although we are mostly interested in systems where relativity is important, an improved method to solve the nonrelativistic case is of interest by itself, because it can be applied directly to heavy quarkonia. 

The paper's main result is that the momentum-space equation for the linear potential can be rewritten in such a way that all singularities are removed. The resulting equation can then be solved much more easily using standard numerical methods.
In Sec.~\ref{sec:linpot} we review the treatment of the linear potential in momentum space and the resulting form of the Schr\"odinger equation with a singular kernel. A proof that the singularity is of Cauchy principal value type and which does not use a partial wave decomposition is given in Appendix \ref{Sec:PV}. In Sec.~\ref{sec:nonsingular} we show how the principal value singularity can be removed, such that the resulting equation has a singularity-free kernel in all partial waves. We present in Sec.~\ref{sec:results} numerical results obtained with the singularity-free equation and demonstrate their numerical accuracy and stability. In Sec.~\ref{sec:summary} we summarize our findings and draw our conclusions.

\section{Linear confinement potential in momentum space}
\label{sec:linpot}
\noindent
The linear potential in coordinate space,
\begin{equation}
\tilde{V}({\bf r})=\sigma r \, ,
\label{eq:linr}
\end{equation}
whose slope $\sigma$ is also called the ``string tension'', cannot be Fourier transformed directly to momentum space. 
Instead, one can introduce a screened potential that depends on a screening parameter $\epsilon$ and whose Fourier transform does exist. The momentum-space version of the linear potential is then defined as the unscreened limit of the screened potential's Fourier transform.  A popular choice is
\begin{equation}
\tilde{V}_{S,\epsilon}(\mathbf{r})=\sigma re^{-\epsilon r}=\sigma\cfrac{\partial^{2}}{\partial\epsilon^{2}}\cfrac{e^{-\epsilon r}}{r} \, ,
\end{equation}
whose Fourier transform is obtained most easily from the second derivative of a Yukawa potential,
\begin{align}
V_{S,\epsilon}(\mathbf{q}) & =\int d^{3}re^{i\mathbf{q} \cdot \mathbf{r}}\tilde{V}_{S,\epsilon}(\mathbf{r})=\sigma\cfrac{\partial^{2}}{\partial\epsilon^{2}}\int d^{3}re^{i\mathbf{q}\cdot \mathbf{r}}\cfrac{e^{-\epsilon r}}{r}
\nonumber\\
 & =-\cfrac{8\pi\sigma}{\left(q^{2}+\epsilon^{2}\right)^{2}}+\cfrac{32\pi\sigma\epsilon^{2}}{\left(q^{2}+\epsilon^{2}\right)^{3}} \, .
 \label{eq:linrscr}
\end{align}
For instance, Maung {\it et al.}\ in Ref.~\cite{Maung:1993aa} perform a partial wave decomposition of the momentum-space Schr\" odinger equation with this screened potential and then take the unscreened limit $\epsilon \rightarrow 0$. 

Eyre and Vary \cite{Eyre:1986} also use the form (\ref{eq:linrscr}), but keep a small non-zero value for $\epsilon$. However, because the screened potential (\ref{eq:linrscr}) has no bound state solutions due to barrier penetration, they subtract a constant $c$ from the potential, leading to $\tilde{V}_\mathrm{EV}({\bf r})=e^{-\epsilon r}(\sigma r - c)$. The constant $c$ is  chosen large enough to support real bound states, and the corresponding binding energies are then ``corrected'' by adding $c$ again.

Gross and Milana \cite{Gross:1991te} and Savkli and Gross \cite{Savkli:1999me} start from (\ref{eq:linrscr}) as well, but after analyzing the behavior of the second term of the Fourier transform they replace it by a Dirac delta function, after which the limit $\epsilon \rightarrow 0$ can be taken.

The latter result can be obtained more directly by choosing as screened potential
\begin{equation}
\tilde{V}_{L,\epsilon}({\bf r})=-\cfrac{\sigma}{\epsilon}\left(e^{-\epsilon r}-1\right) \, ,
\label{eq:VLe}
\end{equation}
 which can also be written
\begin{equation}
\tilde{V}_{L,\epsilon}(\mathbf{r})=\tilde{V}_{A,\epsilon}(\mathbf{r})-\tilde{V}_{A,\epsilon}(0) \, ,
\label{eq:3}
\end{equation}
with 
\begin{equation}
\tilde{V}_{A,\epsilon}(\mathbf{r}) = -\cfrac{\sigma}{\epsilon} e^{-\epsilon r} \, .
\end{equation}
It is then clear that when going to momentum space a delta function will arise from the constant term $\tilde{V}_{A,\epsilon}(0)=-\sigma/\epsilon$. 

The momentum-space form of the screened linear potential (\ref{eq:VLe}) is obtained as
\begin{align}
V_{L,\epsilon}({\bf q}) 
 & =\int d^{3}r\left[\tilde{V}_{A,\epsilon}(\mathbf{r})-\tilde{V}_{A,\epsilon}(0)\right]e^{i\mathbf{q}\cdot \mathbf{r}}
  \nonumber\\
 & =\int d^{3}r\tilde{V}_{A,\epsilon}(\mathbf{r})e^{i\mathbf{q}\cdot \mathbf{r}}-(2\pi)^{3}\delta^{(3)}(\mathbf{q})\tilde{V}_{A,\epsilon}(0) 
 \nonumber \\
 & =V_{A,\epsilon}({\bf q})-(2\pi)^{3}\delta^{(3)}(\mathbf{q})\int\cfrac{d^{3}q'}{(2\pi)^{3}}V_{A,\epsilon}({\bf q}') \, ,
\label{eq:mom}
\end{align}
and 
\begin{equation}
V_{A,\epsilon}(\mathbf{q})=-\cfrac{8\pi\sigma}{\left(q^{2}+\epsilon^{2}\right)^{2}} \, .
\label{eq:V_Ae}
\end{equation}

Before going to the unscreened limit, it should be mentioned that the potential (\ref{eq:VLe}) may be interesting also for finite $\epsilon$. It has been argued that the effect of string breaking could be simulated to some extent with a potential that rises almost linearly only up to a certain distance and then turns flat, which is exactly the behavior of (\ref{eq:VLe}). Therefore it will also be of interest to study its solutions for varying values of $\epsilon$.

If one takes now the limit $\epsilon\rightarrow0$, one gets a potential
that is singular at $\mathbf{q}=0,$ but  has a ``built-in'' subtraction term
that regularizes integrations over the singularity:
\begin{align}
V_{L}({\bf q}) & =\lim_{\epsilon\rightarrow0}\left[V_{A,\epsilon}({\bf q})-(2\pi)^{3}\delta^{(3)}({\bf q})\int\cfrac{d^{3}q'}{(2\pi)^{3}}V_{A,\epsilon}({\bf q}')\right]
\nonumber\\
 & =V_{A}({\bf q})-(2\pi)^{3}\delta^{(3)}(\mathbf{q})\int\cfrac{d^{3}q'}{(2\pi)^{3}}V_{A}({\bf q}'). 
 \label{eq:VL}
 \end{align}
As a check, it is easy to calculate the  Fourier transform of $V_L$ back into $r$-space at the point $r=0$,
\begin{equation}
\int\cfrac{d^{3}q}{\left(2\pi\right)^{3}}V_{L}(\mathbf{q})=0 \, .
\label{eq:intVL}
\end{equation}
The linear potential (\ref{eq:linr}) vanishes at $r=0$, which is correctly reproduced by (\ref{eq:VL}).

With the form (\ref{eq:VL}) of the linear potential $V_L({\bf q})$, the Schr\" odinger equation for a two-body system with reduced mass $m_R$ becomes

\begin{equation}
\frac{p^2}{2m_R}  \Psi( {\bf p}) 
+  
\mathrm {P}\!\!\!\int  \frac{d^3 k}{(2\pi)^3}  \, V_A({\bf p}-{\bf k})
\left[  \Psi( {\bf k})  
 - \Psi( {\bf p})  \right] =
 E  \Psi( {\bf p}) \, ,
 \label{eq:SE}
\end{equation}
or, more explicitly,
\begin{equation}
\frac{p^2}{2m_R}  \Psi( {\bf p}) 
-8\pi\sigma  \,
\mathrm {P}\!\!\! \int  \frac{d^3 k}{(2\pi)^3}  
\frac{  \Psi( {\bf k})  
 - \Psi( {\bf p})  }{({\bf p}-{\bf k})^4}
  =
 E  \Psi( {\bf p}) \, .
 \label{eq:SEa}
\end{equation}
The strong singularity in the kernel of (\ref{eq:SEa}) at ${\bf k}={\bf p}$ is weakened by the subtraction term, and together with the definition of $V_L$ in terms of the limit $\epsilon \rightarrow 0$ in Eq.~(\ref{eq:VL}) it reduces to a Cauchy principal value singularity which makes the integral well-defined (for Cauchy principal value integration we use the symbol ``$\mathrm {P}\!\! \int$''). A proof that the singularity is indeed of Cauchy principal value type is given in Appendix \ref{Sec:PV}.

Next we project (\ref{eq:SE}) onto partial wave $\ell$, which leads to the appealingly simple equation
\begin{multline}
\frac{p^2}{2m_R} 
\psi_\ell(p) 
+  
\mathrm {P}\!\!\! \int_0^\infty  \frac{d k \, k^2}{(2\pi)^3}
\left[
V_{A,\ell}(p,k)
\psi_\ell(k)  \right. \\
\left.
- V_{A,0} (p,k)
\psi_\ell(p) 
 \right] 
   =
 E \psi_\ell(p)  \, ,
 \label{eq:2SPW}
\end{multline}
where the subtraction term generated by the delta function in (\ref{eq:VL}) contains only the $S$-wave potential. 
The partial-wave  matrix elements of the potential $V_A$ are
\begin{multline}
V_{A,\ell}(p,k)  =
2\pi (-8\pi\sigma)
\left[ 
\frac{2 P_\ell(y)}{\left( p^2-k^2 \right)^2}
\right. \\
\left.
-
\frac{P'_\ell(y)}{\left( 2p k \right)^2} \ln \left( \frac{p+k}{p-k}\right)^2
+
\frac{2 w'_{\ell-1}(y)}{\left( 2p k \right)^2}
\right] \, ,
\label{eq:2VAPW}
\end{multline}
where 
\begin{equation}
y=\frac{p^2+k^2}{2pk} \, ,
\end{equation}
$P_\ell$ is a Legendre polynomial, $w_{\ell-1}(y)$ is a polynomial of degree $\ell-1$ defined as
\begin{equation}
 w_{\ell-1}(y) \equiv \sum_{m=1}^\ell \frac{1}{m} P_{\ell-m}(y) P_{m-1}(y)  \, , 
\end{equation}
and the prime in $P'_\ell$ and $w'_{\ell-1}$ means a derivative with respect to the argument $y$.
Equations (\ref{eq:2SPW}) and (\ref{eq:2VAPW}) are derived in Appendix \ref{Sec:PW}.

\section{Removal of the singularities in the kernel}
\label{sec:nonsingular}

\noindent
The kernel (\ref{eq:2VAPW}) in the Schr\" odinger equation (\ref{eq:2SPW}) contains singularities. The first term in (\ref{eq:2VAPW}) has a double pole at $k=p$ in all partial waves, but  in (\ref{eq:2SPW})  it  reduces to a principal value singularity. The second term in (\ref{eq:2VAPW}), present in all partial waves with $\ell \ge 1$, diverges logarithmically at $k=p$ and is therefore integrable. The last term involving $w'_{\ell-1}(y)$, which contributes only when $\ell \ge 2$, is not singular at all. So, in principle, equation (\ref{eq:2SPW}) can be solved numerically as it stands.

However, the numerical integration of singular functions requires special care and typically also more computing time. From the practical point of view it would be a considerable advantage to avoid those singularities altogether. 

It has been known for a long time that the logarithmic singularity can be eliminated by a simple subtraction technique due to Land\'e \cite{Lande+Kwon}.  
We will show now that a different subtraction can remove also the principal value singularity.

\subsection{Subtraction of the principal value singularity}

In this section it is important to keep the difference between ordinary and Cauchy principal value integrals explicit in our notation. 

The Schr\" odinger equation (\ref{eq:2SPW}) with the potential $V_A$ of (\ref{eq:2VAPW}) is
\begin{align}
&\frac{p^2}{2m_R} 
\psi_\ell(p) 
 - \frac{2\sigma}{\pi}  
\mathrm {P}\!\!\! \int_0^\infty  \!\!\!\! dk 
\biggl\{
\frac{2 k^2}{(k^2-p^2)^2}
\bigl[
P_\ell(y)\psi_\ell(k)-\psi_\ell(p)  
\bigr]  
\nonumber\\ 
&  -
\frac{P'_\ell(y)}{4p^2} \ln \left( \frac{p+k}{p-k} \right)^2 \psi_\ell(k)
+
\frac{2 w'_{\ell-1}(y)}{4p^2} \psi_\ell(k)
\biggr\}
   =
 E \psi_\ell(p)  \, .
 \label{eq:3SPW}
\end{align}
We now turn our attention to the most singular part of the integral in (\ref{eq:3SPW}), namely
\begin{multline}
I_1 \equiv
\mathrm{P}\!\! \int_0^\infty dk 
 \frac{2k^2}{(k^2-p^2)^2}
 \left[ P_\ell(y) \psi_\ell(k)-\psi_\ell(p) \right]  \, .
 \label{eq:I1}
 \end{multline}
Because  $P_\ell(y)=1$ at $k=p$, the numerator vanishes at the singularity in all partial waves, reducing the double pole to a single pole.
To see this, we expand the factor in brackets in a Taylor series around $k=p$:
\begin{align}
  P_\ell &(y)  \psi_\ell(k)-\psi_\ell(p) \nonumber\\
 & =  \psi_\ell(p)  +
 (k-p) \left[  P'_\ell(y) \frac{\partial y}{\partial k} \psi_\ell(k)+
P_\ell(y)\psi'_\ell(k)\right]_{k=p}
 \nonumber\\
& \qquad +(k-p)^2 R_\ell(k)- 
\psi_\ell(p) 
 \nonumber\\
& = (k-p) \psi'_\ell(p) + (k-p)^2 R_\ell(k) \, ,
\end{align}
where we have used that  $\frac{\partial y}{\partial k}|_{k=p}=0$. The function $(k-p)^2R_\ell(k)$ is the remainder of the Taylor series of $P_\ell(y)\psi_\ell(k)$ around $k=p$ once the constant and the term linear in $(k-p)$ have been subtracted. The only relevant property of $R_\ell(k)$ in this context is that it is finite at $k=p$.

The integrand of (\ref{eq:I1}) can therefore be written
\begin{multline}
 \frac{2k^2}{(k^2-p^2)^2}
 \left[ P_\ell(y) \psi_\ell(k)-\psi_\ell(p) \right] \\
 =
 \frac{2k^2}{(k+p)^2} \frac{ \psi'_\ell(p)}{k-p} 
 +  \frac{2k^2 R_\ell(k)}{(k+p)^2} \, ,
\label{eq:taylor1}
\end{multline}
where the singular pole term has now been isolated. It is, however, more useful to further rewrite this expression as
\begin{multline}
 \frac{2k^2}{(k^2-p^2)^2} \left[ P_\ell(y) \psi_\ell(k)-\psi_\ell(p) \right] 
 \\
 =  \frac{p\psi'_\ell(p)}{k^2-p^2} +\psi'_\ell(p)\frac{2k+p}{(k+p)^2} +\frac{2k^2 R_\ell(k)}{(k+p)^2} \, .
\end{multline}
This form of the integrand has the advantage over (\ref{eq:taylor1}) that its singular term can be integrated analytically. 
Using it as a subtraction term, we write the principal value integral $I_1$ as an ordinary integral over a now non-singular integrand plus a principal value integral that can be calculated analytically:
\begin{align}
I_1& = \mathrm{P}\!\! \int_0^\infty dk 
 \frac{2k^2}{(k^2-p^2)^2}
 \bigl[ P_\ell(y) \psi_\ell(k)-\psi_\ell(p) \bigr] \nonumber\\
& = \int_0^\infty dk \Bigl\{
 \frac{2k^2}{(k^2-p^2)^2}
 \left[ P_\ell(y) \psi_\ell(k)-
\psi_\ell(p) \right] 
\nonumber \\
& \qquad\qquad - \frac{p\psi'_\ell(p)}{k^2-p^2} \Bigr\}
 + 
 p\psi'_\ell(p) \mathrm{P}\!\! \int_0^\infty  \frac{dk}{k^2-p^2} \, .
\end{align}
In this case, the principal value integration can be performed very easily,
\begin{equation}
 \mathrm{P}\!\!  \int_0^\infty \frac{dk}{k^2-p^2}=0  \, ,
\end{equation}
and we arrive at the simple result
\begin{equation}
I_1 = 
\int_0^\infty \!\! dk \Biggl\{
 \frac{2k^2}{(k^2-p^2)^2}
 \left[ P_\ell(y) \psi_\ell(k)-
\psi_\ell(p) \right] 
- \frac{p\psi'_\ell(p)}{k^2-p^2} \Biggr\}   \, .
   \label{eq:I1reg}
\end{equation}
The price to pay for this simplification is that the derivative of the wave function appears in the integrand. However, this is no significant complication if the method of solving the integral equation uses an expansion of $\psi_\ell(p)$ into a set of basis functions whose derivatives can be easily calculated.

\subsection{Subtraction of the logarithmic singularity}
\label{sec:logsub}

\noindent
The second singular integrand of (\ref{eq:3SPW})  is
\begin{multline}
I_2 \equiv
 -\frac{1}{4p^2}
 \int_0^\infty dk 
 \ln\left( \frac{p+k}{p-k}\right)^2
 P'_\ell(y)\psi_\ell(k)
\\
 = -\frac{1}{2p^2}
 \int_0^\infty dk \, 
Q_0(y)
 P'_\ell(y)\psi_\ell(k) \, ,
 \label{eq:I2def}
\end{multline}
where $Q_\ell$ are the Legendre functions of the second kind. 
In this case, we can take advantage of the known result \cite{Kwon:1978}
\begin{equation}
\int_0^\infty dk \frac{Q_0(y)}{k}=\frac{\pi^2}{2} 
\label{eq:intQ0y}
\end{equation}
to bring $I_2$ into the following form:
\begin{multline}
I_2  =
 -\frac{1}{2p^2}
 \int_0^\infty dk \, 
Q_0(y)
\left[
 P'_\ell(y)\psi_\ell(k) 
 -\frac{p}{k} P'_\ell(1) \psi_\ell(p) \right]
\\
 -\frac{1}{2p^2} p  P'_\ell(1) \psi_\ell(p) 
 \int_0^\infty dk \, 
 \frac{Q_0(y)}{k}  
 \\
  =
 -\frac{1}{2p^2}
 \int_0^\infty dk \, 
Q_0(y)
\left[
 P'_\ell(y)\psi_\ell(k) 
 -\frac{p}{k} P'_\ell(1) \psi_\ell(p) \right]
\\
- \frac{\pi^2 }{4p} P'_\ell(1) \psi_\ell(p) \, .
\label{eq:I2reg}
\end{multline}

It is easy to see that the factor in brackets in the integrand is proportional to $(k-p)$ near $k=p$, such that $(k-p)Q_0(y)$ vanishes at that point. The subtracted integrand is therefore no longer singular. The derivatives of the Legendre polynomials at $y=1$ can be calculated from the well-known relation
\begin{equation}
P'_\ell(1)=\frac{\ell (\ell +1)}{2} \, .
\end{equation}

Substitution of the results (\ref{eq:I1reg}) and (\ref{eq:I2reg}) into the partial wave Schr\"odinger equation (\ref{eq:3SPW}) gives us the final result
\begin{widetext}
\begin{multline}
\left[
\frac{p^2}{2m_R} + \frac{\sigma \pi }{2p} P'_\ell(1) \right] \psi_\ell(p)
- 
 \frac{2\sigma}{\pi}\int_0^\infty dk 
 \Biggl\{ 
  \left[  \frac{2k^2}{\left(k^2-p^2\right)^2}   \Big( P_\ell(y) \psi_\ell(k)- \psi_\ell(p) \Big)  -\frac{p \psi'_\ell(p)}{k^2-p^2} \right]
\\
-\frac{1}{4p^2} \ln \left( \frac{p+k}{p-k} \right)^2 
 \left[ P'_\ell(y) \psi_\ell(k) - P'_\ell(1)\frac{p}{k}\psi_\ell(p) 
 \right]
  +\frac{w'_{\ell-1}(y)}{2p^2} \psi_\ell(k)
 \Biggr\}
 = E \psi_\ell(p) \, .
\label{eq:PWSEsub}
\end{multline}
\end{widetext}
In this equation, the principal value singularity and the logarithmic singularity in the kernel have both been removed {\it in all partial waves}, and the integrand is a smooth function at the originally singular point $k=p$. It is therefore much easier to solve numerically than the original singular equation (\ref{eq:3SPW}) if one is able to supply the derivative $\psi'_\ell(p)$ of the unknown wave function in the new subtraction term. This is easy when the chosen method to solve the integral equation (\ref{eq:PWSEsub}) numerically is to expand the wave function in a set of appropriate basis functions (i.e., a Galerkin method). For instance, in this work we used a basis of cubic spline functions, modified to satisfy the correct boundary conditions. It is less easy in collocation methods, where one demands the equation to hold exactly at a certain set of collocation points, which are usually the points associated with some quadrature rule.  

In fact, after deriving Eq.~(\ref{eq:PWSEsub}) and convincing ourselves of its advantages over (\ref{eq:3SPW}), we found that  Deloff \cite{Deloff:2007} had already written down the $S$-wave version of the subtracted equation. However, Ref.~\cite{Deloff:2007} did not pursue it further because it was considered not suitable for the approach proposed there.

\subsection{The screened linear potential for finite screening parameter}
\label{sec:SLP}
From the results we have obtained so far, it is easy to derive the partial wave Schr\"odinger equation for the screened linear potential Eq.~(\ref{eq:VLe}) in momentum space. 
We can write the partial-wave matrix elements of the screened linear potential (\ref{eq:V_Ae})
\begin{equation}
\langle p \, \ell m | V_{A,\epsilon} | k \,  \ell m \rangle =
2\pi  \frac{(-8\pi\sigma)}{(2pk)^2}\int_{-1}^1 dx \frac{P_\ell(x)}{(y_\epsilon- x)^2} \, ,
\end{equation}
with 
\begin{equation}
y_\epsilon = \frac{k^2+p^2+\epsilon^2}{2pk} \, .
\end{equation}
Compared to the unscreened case (\ref{eq:VAPW}), the only modification necessary is therefore to replace $y$ by $y_\epsilon$.
The result is
\begin{align}
& \langle p \, \ell m | V_{A,\epsilon} | k \, \ell m \rangle 
\nonumber\\
& = 2\pi 
\frac{(-8\pi\sigma)}{(2pk)^2}(-2)
\bigl[  P_\ell(y_\epsilon) Q'_0(y_\epsilon)+ P'_\ell(y_\epsilon) Q_0(y_\epsilon) 
\nonumber\\
 & 
\qquad \qquad \qquad \qquad  -w'_{\ell-1}(y_\epsilon) \bigr] 
\nonumber \\
& = 2\pi (-8\pi\sigma)
\Biggl\{
\frac{2 P_\ell(y_\epsilon)}{\left[ ( p-k )^2+\epsilon^2\right] \left[ ( p+k )^2+\epsilon^2\right] }
\nonumber\\
&\qquad 
-
\frac{P'_\ell(y_\epsilon)}{\left( 2p k \right)^2} \ln \left[ \frac{(p+k)^2+\epsilon^2}{(p-k)^2+\epsilon^2}\right]
+
\frac{2 w'_{\ell-1}(y_\epsilon)}{\left( 2p k \right)^2}
\Biggr\} .
\label{eq:VAePW}
\end{align}
As long as $\epsilon$ remains finite, this potential is not singular and can be used in the Schr\"odinger equation without further modifications. However, when $\epsilon$ becomes very small it becomes ``almost singular'' and therefore numerically very difficult to control. We found that a subtraction of the log-term makes the numerical solution of the Schr\"odinger equation converge significantly faster and the results more stable. 

For this purpose, a generalization of Eq.~(\ref{eq:intQ0y}),
\begin{equation}
\int_0^\infty dk \,  \frac{Q_0(y_\epsilon)}{k}=\frac{\pi^2}{2}-\pi\arctan\frac{\epsilon}{p} \, ,
\end{equation}
is used to write 
\begin{align}
\frac{\sigma}{\pi p^2} & \int_0^\infty dk \, P'_\ell(y_\epsilon) Q_0(y_\epsilon) \psi_\ell(k) \nonumber \\
& =
\frac{\sigma}{\pi p^2} \int_0^\infty dk \, Q_0(y_\epsilon)
\left[ P'_\ell(y_\epsilon)  \psi_\ell(k) 
- \frac{p}{k} P'_\ell(\bar{y}_\epsilon)  \psi_\ell(p) \right] \nonumber \\
& \qquad  + \frac{\sigma}{p} P'_\ell(\bar{y}_\epsilon) \left(\frac{\pi}{2}-\arctan \frac{\epsilon}{p} \right) \psi_\ell(p)
\, ,
\end{align}
where $\bar{y}_\epsilon = 1+\epsilon^2/2p^2$ is the value of $y_\epsilon$ at the point $k=p$.

The Schr\"odinger equation for the screened linear potential of Eq.~(\ref{eq:VLe}) with finite screening parameter $\epsilon$ can then be written in momentum space

\begin{multline}
\left[
\frac{p^2}{2m_R} + \frac{\sigma}{p} P'_\ell(\bar{y}_\epsilon) \left(\frac{\pi}{2}-\arctan \frac{\epsilon}{p} \right) \right] \psi_\ell(p)
\\
- 
 \frac{2\sigma}{\pi}\int_0^\infty dk 
 \Biggl\{ 
  \frac{2k^2}{\left[ ( p-k )^2+\epsilon^2\right] \left[ ( p+k )^2+\epsilon^2\right]}  
\\
\times   \Big[ P_\ell(y_\epsilon) \psi_\ell(k)- \psi_\ell(p) \Big]  
-\frac{1}{4p^2} 
\ln \left[ \frac{(p+k)^2+\epsilon^2}{(p-k)^2+\epsilon^2}\right] 
\\
\times
 \left[ P'_\ell(y_\epsilon) \psi_\ell(k) - P'_\ell(\bar{y}_\epsilon)\frac{p}{k}\psi_\ell(p) 
 \right]
  +\frac{w'_{\ell-1}(y_\epsilon)}{2p^2} \psi_\ell(k)
 \Biggr\}
 \\
 = E \psi_\ell(p) \, .
\label{eq:PWSEepssub}
\end{multline}

\subsection{Addition of a Coulomb-type potential}
\label{sec:Coulomb}

The often used Cornell-type potentials combine a linear with a Coulomb potential, which in coordinate space reads
\begin{equation}
\tilde{V}_C({\bf r})= -\frac{\alpha}{r} \, .
\end{equation}
Its Fourier-transform is well known,
\begin{equation}
V_{C}({\bf q}) 
 =\int d^{3}r \tilde{V}_C({\bf r}) e^{i{\bf q}\cdot {\bf r}}
 =-\frac{4\pi\alpha}{{\bf q}^2} \, ,
\end{equation}
and the partial wave matrix elements are
\begin{multline}
\langle p \, \ell m | V_C | k \,  \ell m \rangle =
2\pi  \frac{(-4\pi\alpha)}{2pk}\int_{-1}^1 dx \frac{P_\ell(x)}{y- x}  \\
=-\frac{8\pi^2 \alpha}{pk} Q_\ell(y)
= -\frac{8\pi^2 \alpha}{pk} \left[P_\ell(y)Q_0(y)-w_{\ell-1}(y)\right]
\, .
\end{multline}
The last form makes it obvious that the singularity at $k=p$ is of the same kind in all partial waves. We encountered it already in Eq.~(\ref{eq:I2def}) as one of the singular terms in the linear potential. In the kernel of the Schr\"odinger equation, the singularity can therefore be subtracted with the same technique \cite{Kwon:1978,Norbury:1994zp}, namely 
\begin{multline}
 -\frac{8\pi^2 \alpha}{p }
 \int_0^\infty  \frac{dk \, k}{(2\pi)^3}
 P_\ell(y) Q_0(y)
\psi_\ell(k) 
=
-\frac{\alpha \pi}{2}p \psi_\ell(p)  \\
-\frac{\alpha}{ \pi}
\int_0^\infty dk \, Q_0(y)
\left[ \frac{k}{p} P_\ell(y) \psi_\ell(k) - \frac{p}{k} \psi_\ell(p) \right] \, ,
\label{eq:Creg}
\end{multline}
where $P_\ell(1)=1$ was used.

The resulting singularity-free version of the Schr\"odinger equation for the unscreened linear plus Coulomb potential in momentum space is
\begin{widetext}
\begin{multline}
\left[
\frac{p^2}{2m_R} + \frac{\sigma \pi }{2p} P'_\ell(1) -\frac{\alpha \pi p}{2}\right] \psi_\ell(p)
- 
 \frac{2\sigma}{\pi}\int_0^\infty dk 
 \Biggl\{ 
  \left[  \frac{2k^2}{\left(k^2-p^2\right)^2}   \Big( P_\ell(y) \psi_\ell(k)- \psi_\ell(p) \Big)  -\frac{p \psi'_\ell(p)}{k^2-p^2} \right]
\\
-\frac{1}{4p^2} \ln \left( \frac{p+k}{p-k} \right)^2 
 \left[ P'_\ell(y) \psi_\ell(k) - P'_\ell(1)\frac{p}{k}\psi_\ell(p) 
 \right]
  +\frac{w'_{\ell-1}(y)}{2p^2} \psi_\ell(k)
 \Biggr\}
 \\
 -\frac{\alpha}{\pi}
 \int_0^\infty dk 
 \Biggl\{ \frac12
  \ln \left( \frac{p+k}{p-k} \right)^2
   \left[ \frac{k}{p} P_\ell(y) \psi_\ell(k) - \frac{p}{k}\psi_\ell(p) 
 \right]
  -\frac{k}{p} w_{\ell-1}(y)\psi_\ell(k)
 \Biggr\}
 = E \psi_\ell(p) \, .
\label{eq:PWSECsub}
\end{multline}
\end{widetext}
It is an easy task to adapt the results of this section for the case of an exponentially screened Coulomb potential. This can be done in close analogy with the derivation of the screened linear potential shown in Sec.~\ref{sec:SLP}.  

\section{Numerical results}
\label{sec:results}
In this section we present the numerical results obtained with the singularity-free equation introduced in the Sec.~\ref{sec:nonsingular}. 

\subsection{Expansion into splines}
\label{sec:splines}
We solved the momentum-space Schr\"odinger equation numerically by expanding the wave function in the basis of cubic B-splines described in detail in Refs.~\cite{Gross:1994he,Uzz99} and thereby converting the integral equation into a generalized eigenvalue problem. The eigenvalues are the binding energies of the system, whereas the corresponding eigenvectors contain the spline expansion coefficients. 

The original basis of $N$ cubic B-splines $b_i(x)$ is defined for $x \in [0,1]$ with equidistant knots. To construct a basis for functions  of momenta $p \in [0,\infty[$, we use a map 
\begin{equation}
p(x)=\Lambda \tan \frac{\pi x}{2} \, ,
\end{equation}
where $\Lambda$ is a scale parameter,
and define our basis functions as
\begin{equation}
\beta^\ell_i(p) = \left( \frac{p}{E_p} \right)^\ell b_i \left( \frac{2}{\pi}\arctan \frac{p}{\Lambda} \right) \, ,
\end{equation}
with $E_p = \sqrt{m^2+p^2}$;  $m$ is chosen as a particle mass for simplicity, although here it plays the role of a free parameter. For small momenta, the wave function in partial wave $\ell$ behaves like  $\psi_\ell(p) \sim p^\ell$. Because the first spline does not vanish at $p=0$, the factor $(p/E_p)^\ell$ is a simple way to make sure that the basis functions are compatible with this constraint \cite{Gross:1994he}. However, no particular significance is connected with the appearance of a term that resembles a relativistic energy. Instead, one can use just as well, for instance, a factor $[p/(m+p)]^\ell$. In the calculations of this paper we use $\Lambda=m=1$ in units of $(2m_R \sigma)^{1/3}$.

The wave function is expanded in the spline basis,
\begin{equation}
\psi_\ell(p)=\sum_{j=1}^N c_j \beta^\ell_j(p) \, ,
\end{equation}
and the partial wave Schr\"odinger equation (for a  total potential $V_\ell(p,k)$) is then multiplied by $p^2 \beta^\ell_i(p)$ and integrated over $p$. The result is a generalized eigenvalue equation for the expansion coefficients $c_j$ of the form
\begin{equation}
\sum_{j}(A_{ij}+V_{ij}) c_j = E \sum_{j} B_{ij}c_j \, ,
\end{equation}
where the matrices are defined as 
\begin{align}
A_{ij} & =  \int_0^\infty dp \,  p^2 \beta^\ell_i(p)\beta^\ell_j(p)  \frac{p^2}{2m_R} \, ,\nonumber\\
B_{ij} & =  \int_0^\infty dp \,  p^2  \beta^\ell_i(p)\beta^\ell_j(p) \, ,\nonumber\\
V_{ij} & =  \int_0^\infty dp \,  p^2  \beta^\ell_i(p)\int_0^\infty  \frac{d k \, k^2}{(2\pi)^3} \,  V_\ell(p,k) \beta^\ell_j(k) \, .
\end{align}
This generic form of the potential matrix has to be adapted to the singularity-free form according to Eqs.~(\ref{eq:PWSEsub}), (\ref{eq:PWSEepssub}),  or (\ref{eq:PWSECsub}), for the singular potentials discussed in the previous sections.

\begingroup
\begin{table*}[tb]
\centering
\caption{The ten lowest energy eigenvalues $E_n$ of the unscreened linear potential with $\ell=0$ (S-wave), obtained by solving Eq.~(\ref{eq:PWSEsub}) with an increasing number of splines, $N$, in the  B-spline basis. The last column shows the exact solutions from Eq.~(\ref{eq:AiryEn}). The energies are in units of  $(\sigma^2/2m_R)^{1/3}$.}
\begin{ruledtabular}
\begin{tabular}{crrrrrrrr}
 \text{n} & \multicolumn{1}{c}\textrm{N=12} & \multicolumn{1}{c}\textrm{N=16} & \multicolumn{1}{c}\textrm{N=20} &
  \multicolumn{1}{c}\textrm{N=24} & \multicolumn{1}{c}\textrm{N=36} & \multicolumn{1}{c}\textrm{N=48} & \multicolumn{1}{c}\textrm{N=64} &  \multicolumn{1}{c}\textrm{Exact} \\
\hline
 1 & 2.338121 & 2.338108 & 2.338108 & 2.338107 & 2.338107 &
   2.338107 & 2.338108 & 2.338107 \\
 2 & 4.088498 & 4.087976 & 4.087953 & 4.087950 & 4.087947 &
   4.087949 & 4.087949 & 4.087949 \\
 3 & 5.527017 & 5.520928 & 5.520601 & 5.520568 & 5.520559 &
   5.520559 & 5.520560 & 5.520560 \\
 4 & 6.794183 & 6.788208 & 6.787047 & 6.786787 & 6.786710 &
   6.786707 & 6.786708 & 6.786708 \\
 5 & 8.002342 & 7.956598 & 7.947220 & 7.944767 & 7.944146 &
   7.944135 & 7.944134 & 7.944134 \\
 6 & 9.626868 & 9.156258 & 9.046241 & 9.026388 & 9.022727 &
   9.022657 & 9.022651 & 9.022651 \\
 7 & 11.435079 & 10.273394 & 10.083415 & 10.048670 & 10.040511 &
   10.040201 & 10.040177 & 10.040174 \\
 8 & 12.099834 & 11.147565 & 11.027556 & 11.028855 & 11.009868 &
   11.008626 & 11.008534 & 11.008524 \\
 9 & 14.993451 & 12.941736 & 12.318324 & 12.105283 & 11.940068 &
   11.936344 & 11.936044 & 11.936016 \\
 10 & 19.122419 & 15.309248 & 13.997541 & 13.138047 & 12.839002 &
   12.829770 & 12.828860 & 12.828777 \\
\end{tabular}
\label{tab:SNvar-l0}
\end{ruledtabular}
\end{table*}
\endgroup
\subsection{Results for the linear potential}

\begin{figure}[tb]
\begin{center}
\includegraphics[width=0.45\textwidth]{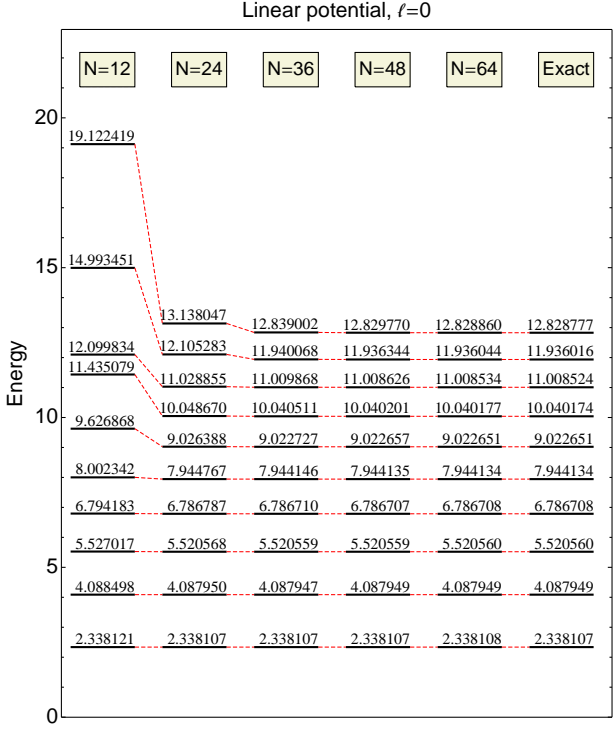}
\caption{(Color online) The lowest 10 energy eigenvalues $E_n$ of the unscreened linear potential with $\ell=0$ (S-wave). The convergence of the eigenvalues  with increasing number of splines, $N$, in the  B-spline basis is shown. The energies are in units of  $(\sigma^2/2m_R)^{1/3}$.}
\label{fig:SNvar-l0}
\end{center}
\end{figure}

\begin{figure}[tb]
\begin{center}
\includegraphics[width=0.5\textwidth]{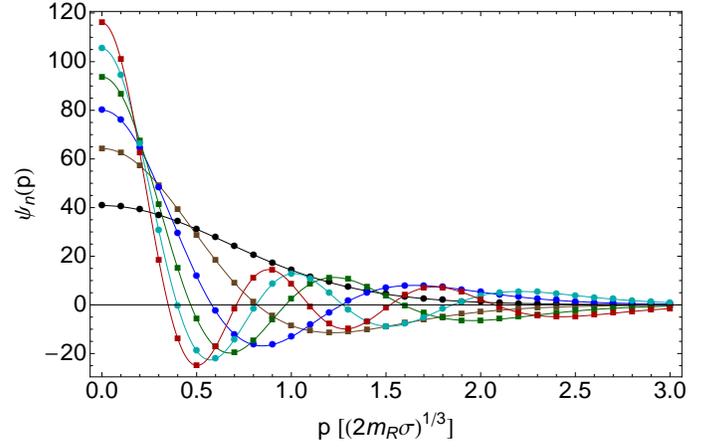}
\caption{(Color online) The solid lines are the S-wave momentum-space wave functions (in arbitrary units) for the  6 lowest eigenstates, calculated in a basis of 64 splines. At $p=0$, the order of the states is from $n=1$ (lowest line) to $n=6$ (highest line). The symbols on the lines represent the exact solutions of Eq.~(\ref{eq:Airyu}), numerically Fourier transformed from coordinate space to momentum space.}
\label{fig:WF-FT}
\end{center}
\end{figure}

Using the expansion into a basis of $N$ cubic B-splines, we solved the singularity-free form of the Schr\"odinger equation, Eq.~(\ref{eq:PWSEsub}), with a linear potential in momentum space. 

The Hamiltonian contains two parameters, the slope $\sigma$ of the linear potential and the reduced mass $m_R$ of the system. It is well known---and can also be derived quite easily from Eq.~(\ref{eq:2SPW})---that the energy eigenvalues scale with $(\sigma^2/2m_R)^{1/3}$. It is therefore sufficient to solve Eq.~(\ref{eq:PWSEsub}) for $\sigma=2 m_R=1$. 

The S-wave equation is of particular interest, because the exact solution in coordinate space is known in terms of the Airy function Ai:
\begin{equation}
E^{\ell=0}_n=-z_n \left( \frac{\sigma^2}{2m_R} \right)^{1/3} \, , \mbox{ with  } \mathrm{Ai}(z_n)=0 \, ,
\label{eq:AiryEn}
\end{equation}
i.e., $z_n$ is the n-th root of the Airy function $\mathrm{Ai}(z)$. Notice that $z_n$ is negative for all $n$. If the coordinate-space eigenstate wave functions are written 
\begin{equation}
\Psi_{n\ell m}({\bf r})=\frac{u_{n\ell}(r)}{r}Y_{\ell m}(\hat{\bf r}) \, ,
\end{equation}
the exact S-wave solutions of the radial wave functions are
\begin{equation}
u_{n0}(r)= a_n \mathrm{Ai}[(2m_R\sigma)^{1/3} r +z_n] \, ,
\label{eq:Airyu}
\end{equation}
where the coefficients $a_n$ are determined through the normalization condition
\begin{equation}
\int dr |u_{n\ell}(r)|^2 =1 \, .
\end{equation}
The S-wave is therefore the ideal case to test our numerical methods.

First we investigate the numerical convergence of the energy eigenvalues and the corresponding wave functions as the number of basis splines $N$ increases. Table \ref{tab:SNvar-l0} and Fig.~\ref{fig:SNvar-l0} show that our numerical S-wave energies converge quickly and smoothly to the exact solutions. For the first few excited states, a small number of splines of the order of 20 is already sufficient to obtain very accurate results. For higher radial excitations, or if more accuracy is required, the spline basis may be increased as needed.

\begin{table}[tb]
\centering
\caption{Energy eigenvalues $E_n$ of the unscreened linear potential with $\ell=1$ (P-wave). The convergence of the eigenvalues  with increasing number of splines, $N$, in the  B-spline basis is shown. The energies are in units of  $(\sigma^2/2m_R)^{1/3}$.}
\begin{ruledtabular}
\begin{tabular}{crrrrr}
 \text{n} & \multicolumn{1}{c}\textrm{N=12} & 
   \multicolumn{1}{c}\textrm{N=24} & \multicolumn{1}{c}\textrm{N=36} & \multicolumn{1}{c}\textrm{N=48} & \multicolumn{1}{c}\textrm{N=64} \\
\hline
 1 & 3.361313 & 3.361257 & 3.361258 &
   3.361257 & 3.361258 \\
 2 & 4.886359 &  4.884455 & 4.884453 &
   4.884452 & 4.884455 \\
 3 & 6.220271 &  6.207647 & 6.207623 &
   6.207621 & 6.207626 \\
 4 & 7.407860 &  7.405868 & 7.405667 &
   7.405661 & 7.405667 \\
 5 & 8.702469 &  8.516776 & 8.515259 &
   8.515230 & 8.515235 \\
 6 & 10.554019 &  9.564306 & 9.557759 &
   9.557619 & 9.557617 \\
 7 & 11.809212 &  10.555695 & 10.547168 &
   10.546563 & 10.546526 \\
 8 & 13.208354 &  11.549231 & 11.493800 &
   11.491595 & 11.491441 \\
 9 & 16.518592 &  12.692407 & 12.405350 &
   12.399775 & 12.399263 \\
 10 & 21.192198 &  13.446955 & 13.293692 &
   13.276712 & 13.275225 \\
\end{tabular}
\label{tab:SNvar-l1}
\end{ruledtabular}
\end{table}

\begin{table}[tb]
\centering
\caption{Energy eigenvalues $E_n$ of the unscreened linear potential with $\ell=2$ (D-wave). The convergence of the eigenvalues  with increasing number of splines, $N$, in the  B-spline basis is shown. The energies are in units of  $(\sigma^2/2m_R)^{1/3}$.}
\begin{ruledtabular}
\begin{tabular}{crrrrr}
 \text{n} & \multicolumn{1}{c}\textrm{N=12} & 
   \multicolumn{1}{c}\textrm{N=24} & \multicolumn{1}{c}\textrm{N=36} & \multicolumn{1}{c}\textrm{N=48} & \multicolumn{1}{c}\textrm{N=64} \\
\hline
 1 & 4.248388 &  4.248175 & 4.248183 &
   4.248185 & 4.248187 \\
 2 & 5.634684 &  5.629695 & 5.629706 &
   5.629708 & 5.629714 \\
 3 & 6.886542 &  6.868909 & 6.868878 &
   6.868878 & 6.868888 \\
 4 & 8.019971 &  8.010144 & 8.009703 &
   8.009693 & 8.009707 \\
 5 & 9.469902 &  9.080260 & 9.077052 &
   9.076989 & 9.077007 \\
 6 & 11.500265 &  10.095852 & 10.086744 &
   10.086455 & 10.086462 \\
 7 & 12.194657 &  11.061887 & 11.049963 &
   11.048791 & 11.048742 \\
 8 & 14.435939 &  12.104358 & 11.975397 &
   11.971758 & 11.971519 \\
 9 & 18.105465 &  13.226181 & 12.869732 &
   12.861352 & 12.860543 \\
 10 & 23.355617 &  13.814864 & 13.755637 &
   13.722578 & 13.720288 \\
\end{tabular}
\label{tab:SNvar-l2}
\end{ruledtabular}
\end{table}

The S-wave momentum-space wave functions of the six lowest energy eigenstates, calculated in a basis of 64 splines, are shown in Fig.~\ref{fig:WF-FT}. They are compared to the exact $r$-space solutions given in Eq.~(\ref{eq:Airyu}), after they have been numerically Fourier transformed to momentum space. The comparison of the eigenfunctions is a stronger test of our method than the comparison of only the eigenvalues. The agreement is excellent in all cases, which is a clear indication that our numerical momentum-space technique is indeed working very well. 

For partial waves higher than $\ell=0$ no exact solutions are available. Nevertheless, Tables \ref{tab:SNvar-l1} and \ref{tab:SNvar-l2} for the cases of P- and D-waves, respectively, demonstrate that the rate of convergence of the energy eigenvalues is similar to that for S-waves, although it becomes slower with increasing $\ell$ and $n$, as was to be expected. 

\begin{figure}[tb]
\begin{center}
\includegraphics[width=0.45\textwidth]{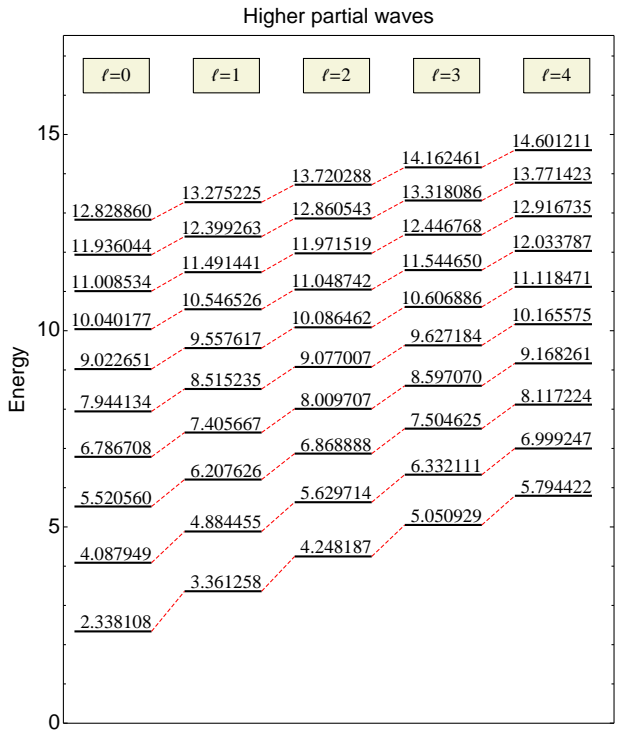}
\caption{(Color online) The ten lowest energy eigenvalues for the linear potential in all partial waves up to $\ell=4$, obtained by solving Eq.~(\ref{eq:PWSEsub}) in a basis of 64 cubic B-splines.  The energies are in units of  $(\sigma^2/2m_R)^{1/3}$.}
\label{fig:Enl0-4}
\end{center}
\end{figure}

We have verified that this trend continues systematically in higher partial waves, and we found no signs of numerical instability. Figure \ref{fig:Enl0-4} shows the lowest ten energy levels in all partial waves from $\ell=0$ to $\ell=4$, calculated with a basis of 64 splines, which gives essentially converged results.

\subsection{Results for the screened linear potential}

We solved Eq.~(\ref{eq:PWSEepssub}) for the screened linear potential of Eq.~(\ref{eq:VLe}) with the numerical method explained in Sec.~\ref{sec:splines}. As already mentioned in Sec.~\ref{sec:linpot}, this potential may be used to simulate the effect of ``string breaking'' when higher excitation energies are reached. However, here we are more interested in the stability of our numerical method. In particular we want to see if the unscreened limit, $\epsilon = 0$, is reached smoothly. 

First we have to verify that the solutions of Eq.~(\ref{eq:PWSEepssub}) converge with an increasing number of basis functions. This is not guaranteed a priori from the success in the unscreened case, because now there is no equivalent of the subtraction of the most singular part in the kernel as in Eq.~(\ref{eq:PWSEsub}). Although the kernel is not strictly singular as long as $\epsilon$ remains finite, for small values it can behave almost as badly as far as the numerical solution is concerned.

Nevertheless, Tab.~\ref{tab:SNvar-l0-e0.01} shows that the energy eigenvalues, here for the case $\epsilon=0.01$, converge very well with  increasing number of splines $N$. We have verified that this good convergence persists also for smaller values of the screening parameter, such as $\epsilon=0.001$ and $\epsilon=0.0001$. 

\begin{table}[tbh]
\centering
\caption{Energy eigenvalues $E_n$ of the screened linear potential with $\ell=0$ (S-wave) with screening parameter $\epsilon=0.01$. The convergence of the eigenvalues  with increasing number of splines, $N$, in the  B-spline basis is shown. The energies are in units of  $(\sigma^2/2m_R)^{1/3}$.}
\begin{ruledtabular}
\begin{tabular}{crrrrr}
 \text{n} & \multicolumn{1}{c}\textrm{N=12} & 
   \multicolumn{1}{c}\textrm{N=24} & \multicolumn{1}{c}\textrm{N=36} & \multicolumn{1}{c}\textrm{N=48} & \multicolumn{1}{c}\textrm{N=64} \\
\hline
 1 & 2.323552 &  2.323540 & 2.323540 &
   2.323540 & 2.323540 \\
 2 & 4.043908 &  4.043439 & 4.043439 &
   4.043439 & 4.043439 \\
 3 & 5.445352 &  5.439425 & 5.439419 &
   5.439418 & 5.439418 \\
 4 & 6.674790 &  6.664184 & 6.664124 &
   6.664123 & 6.664122 \\
 5 & 7.806421 &  7.776688 & 7.776237 &
   7.776227 & 7.776226 \\
 6 & 9.263675 &  8.809037 & 8.806180 &
   8.806129 & 8.806125 \\
 7 & 11.160102 &  9.781259 & 9.772376 &
   9.772155 & 9.772137 \\
 8 & 11.688761 &  10.697552 & 10.687322 &
   10.686448 & 10.686384 \\
 9 & 13.957706 &  11.645487 & 11.560541 &
   11.557625 & 11.557420 \\
 10 & 17.414681 &  12.730431 & 12.398951 &
   12.392183 & 12.391573 \\
\end{tabular}
\label{tab:SNvar-l0-e0.01}
\end{ruledtabular}
\end{table}

Next we consider the behavior of the eigenvalues for smaller and smaller screening parameter. Tables \ref{tab:Screening-l0}, \ref{tab:Screening-l1}, and \ref{tab:Screening-l2}, for the cases of S-, P-, and D-waves, respectively, show that the unscreened limit is reached smoothly. We find no numerical instabilities, such as the ``furcation phenomenon'' reported in Ref.~\cite{Chen:2012}, and we can conclude that the Schr\"odinger equation with both the unscreened and the screened linear potential can be solved in momentum space with very good accuracy and stability. 

\begin{table}[tb]
\centering
\caption{Energy eigenvalues $E_n$ of the screened linear potential with $\ell=0$ (S-wave) with screening parameter $\epsilon$. The last column is the result obtained with the unscreened linear potential. The numerical calculations were performed in a basis with $N=64$ splines. }
\begin{ruledtabular}
\begin{tabular}{crrrrr}
$n$ &\multicolumn{1}{c}{$\epsilon =0.1$} & \multicolumn{1}{c}{$\epsilon=0.01$} & \multicolumn{1}{c}{$\epsilon =0.001$} & \multicolumn{1}{c}{$\epsilon =0.0001$}  & \multicolumn{1}{c}{$\epsilon =0$} \\
  \hline
 1 & 2.193376 & 2.323540 & 2.336650 & 2.337962 & 2.338108 \\
 2 & 3.647650 & 4.043439 & 4.083494 & 4.087504 & 4.087949 \\
 3 & 4.720952 & 5.439418 & 5.512435 & 5.519747 & 5.520560 \\
 4 & 5.582841 & 6.664122 & 6.774429 & 6.785479 & 6.786708 \\
 5 & 6.300437 & 7.776226 & 7.927310 & 7.942449 & 7.944133 \\
 6 & 6.909399 & 8.806125 & 9.000950 & 9.020477 & 9.022651 \\
 7 & 7.431704 & 9.772137 & 10.013305 & 10.037484 & 10.040177 \\
 8 & 7.882197 & 10.686384 & 10.976229 & 11.005298 & 11.008533 \\
 9 & 8.271550 & 11.557420 & 11.898067 & 11.932243 & 11.936044 \\
 10 & 8.607794 & 12.391573 & 12.784988 & 12.824468 & 12.828859 \\
\end{tabular}
\label{tab:Screening-l0}
\end{ruledtabular}
\end{table}

\begin{table}[tb]
\centering
\caption{Energy eigenvalues $E_n$ of the screened linear potential with $\ell=1$ (P-wave) with screening parameter $\epsilon$. The last column is the result obtained with the unscreened linear potential. The numerical calculations were performed in a basis with $N=64$ splines. }
\begin{ruledtabular}
\begin{tabular}{crrrrr}
$n$ &\multicolumn{1}{c}{$\epsilon =0.1$} & \multicolumn{1}{c}{$\epsilon=0.01$} & \multicolumn{1}{c}{$\epsilon =0.001$} & \multicolumn{1}{c}{$\epsilon =0.0001$}  & \multicolumn{1}{c}{$\epsilon =0$} \\
  \hline
 1 & 3.082174 & 3.333193 & 3.358447 & 3.360974 & 3.361258 \\
 2 & 4.272158 & 4.822550 & 4.878256 & 4.883832 & 4.884456 \\
 3 & 5.210944 & 6.106443 & 6.197492 & 6.206609 & 6.207627 \\
 4 & 5.985391 & 7.260964 & 7.391173 & 7.404214 & 7.405669 \\
 5 & 6.639204 & 8.323472 & 8.496024 & 8.513309 & 8.515238 \\
 6 & 7.198401 & 9.315723 & 9.533379 & 9.555186 & 9.557621 \\
 7 & 7.680134 & 10.251774 & 10.516986 & 10.543564 & 10.546531 \\
 8 & 8.096508 & 11.141360 & 11.456350 & 11.487925 & 11.491447 \\
 9 & 8.456507 & 11.991585 & 12.358391 & 12.395170 & 12.399270 \\
 10 & 8.767046 & 12.807839 & 13.228359 & 13.270529 & 13.275234 \\
\end{tabular}
\label{tab:Screening-l1}
\end{ruledtabular}
\end{table}

\begin{table}[tb]
\centering
\caption{Energy eigenvalues $E_n$ of the screened linear potential with $\ell=2$ (D-wave) with screening parameter $\epsilon$. The last column is the result obtained with the unscreened linear potential. The numerical calculations were performed in a basis with $N=64$ splines. }
\begin{ruledtabular}
\begin{tabular}{crrrrr}
$n$ &\multicolumn{1}{c}{$\epsilon =0.1$} & \multicolumn{1}{c}{$\epsilon=0.01$} & \multicolumn{1}{c}{$\epsilon =0.001$} & \multicolumn{1}{c}{$\epsilon =0.0001$}  & \multicolumn{1}{c}{$\epsilon =0$} \\
  \hline
 1 & 3.815719 & 4.204733 & 4.243832 & 4.247747 & 4.248181 \\
 2 & 4.833666 & 5.549257 & 5.621637 & 5.628903 & 5.629705 \\
 3 & 5.666103 & 6.746788 & 6.856605 & 6.867662 & 6.868880 \\
 4 & 6.365415 & 7.842162 & 7.992809 & 8.008030 & 8.009700 \\
 5 & 6.961908 & 8.860760 & 9.055134 & 9.074848 & 9.077000 \\
 6 & 7.475151 & 9.818646 & 10.059296 & 10.083800 & 10.086458 \\
 7 & 7.918745 & 10.726780 & 11.015995 & 11.045563 & 11.048745 \\
 8 & 8.302655 & 11.593057 & 11.932927 & 11.967807 & 11.971529 \\
 9 & 8.634472 & 12.423404 & 12.815868 & 12.856284 & 12.860565 \\
 10 & 8.920167 & 13.222428 & 13.669313 & 13.715463 & 13.720322 \\
\end{tabular}
\label{tab:Screening-l2}
\end{ruledtabular}
\end{table}

\subsection{Linear plus Coulomb potential: fit of the bottomonium spectrum}

To determine whether our numerical framework can be used in practice to perform a least-squares fit to the meson spectrum, we chose the simple Cornell-type potential,
\begin{equation}
\tilde{V}({\bf r})=\sigma r - \frac{\alpha}{r} \, ,
\label{eq:Cornell}
\end{equation}
as our interaction kernel. We have shown in Sec.~\ref{sec:nonsingular} how each of the two components of this potential can be treated conveniently in momentum space by eliminating all singularities. The solutions of the momentum space Schr\"odinger equation depend then on three parameters: the two potential parameters 
$\sigma$ and $\alpha$, and the reduced mass $m_R$. It turned out that the numerical solution of the momentum space Schr\"odinger equation can be done fast enough, such that we were indeed able to determine these three parameters through a least-squares fit to the bottomonium mass spectrum.

The simple potential (\ref{eq:Cornell}) does not contain any spin dependence and consequently is not able to produce a spin splitting of bottomonium states. For our fit we used therefore spin averaged masses, and we limited the fit to states with orbital angular momentum $\ell=0$ or $\ell=1$ below the open flavor threshold. They are marked with an asterisk in Tab.~\ref{tab:masses}.

\begin{table}[tb]
\centering
\caption{Experimental bottomonium masses \cite{PDG:2012} (third column) compared to the theoretical prediction of the model of Eq.~(\ref{eq:Cornell}) (fourth column). The meson masses marked with an asterisk were used in the least-squares fit. When more than one meson is listed in the last column for a given state, their masses have been averaged. The fourth column shows the masses of the states with quantum numbers $n$ and $\ell$ as predicted by the fitted model. All masses are in GeV.}
\begin{ruledtabular}
\begin{tabular}{ccdcr}
$n$ & $\ell$ & \multicolumn{1}{c}\textrm{Experimental mass} & Model mass & Meson(s) \\
  \hline 
 1 & 0 & 9.44298^* & 9.44512 & $\eta_b$(1S), $\Upsilon$(1S) \\
 1 & 1 & 9.89076^* & 9.91265 & $\chi_{b0}$(1P), $\chi_{b1}$(1P) \\
   &    &                  &   & $\chi_{b2}$(1P), $h_b$(1P) \\
 1 & 2 & 10.1637 & 10.1511 &  $\Upsilon$(1D)\\
 2 & 0 & 10.0233^* & 10.0045 & $\Upsilon$ (2S) \\
 2 & 1 & 10.2541^* &  10.2524 & $\chi_{b0}$(2P), $\chi_{b1}$(2P) \\
   &    &                  &   &  $\chi_{b2}$(2P), $h_b$(2P) \\
 3 & 0 & 10.3552^* & 10.3352  & $\Upsilon$ (3S) \\
 3 & 1 & 10.53^* & 10.5244 & $\chi_{b}$(3P)\\
 4 & 0 & 10.5794^* &  10.6014 & $\Upsilon$ (4S) \\
 5 & 0 & 10.876 &  10.8344 & $\Upsilon$ (10860) \\
 6 & 0 & 11.019 &  11.0462 & $\Upsilon$ (11020)\\
\end{tabular}
\label{tab:masses}
\end{ruledtabular}
\end{table}

\begin{figure}[tb]
\begin{center}
\includegraphics[width=0.45\textwidth]{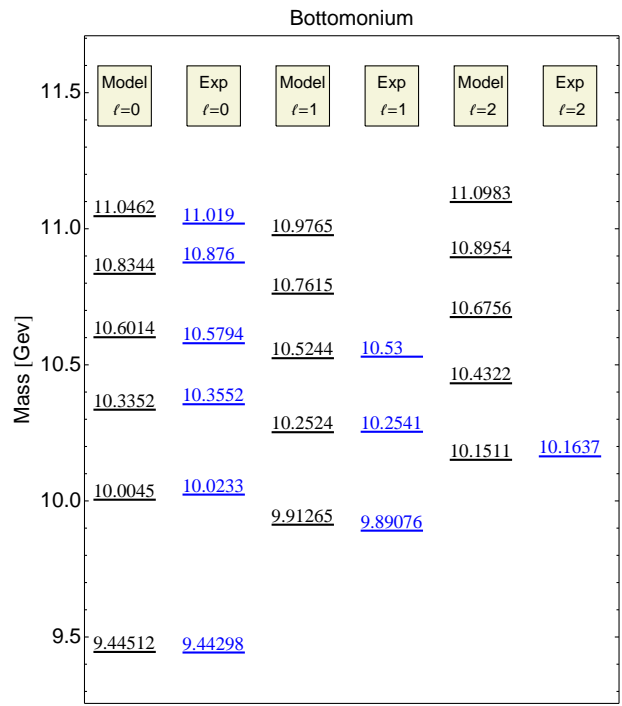}
\caption{(Color online) Bottomonium masses predicted by the model of Eq.~(\ref{eq:Cornell}), compared to the experimental spin-averaged masses. The model parameters were obtained by a least-squares fit to the experimental masses of S- and P-states below open flavor threshold, as indicated in Tab.~\ref{tab:masses}.}
\label{fig:bottomonium}
\end{center}
\end{figure}

The result of the fit is shown in Tab.~\ref{tab:masses} and Fig.~\ref{fig:bottomonium}. The corresponding potential parameters are $\sigma=0.1670$ GeV$^2$ and $\alpha=0.5162$, and the mass of the bottom quark is obtained  as $m_b=4.7931$ GeV from the fitted reduced mass $m_R=m_b/2$. Clearly a good fit is obtained with relatively little effort, and the masses of the three states not used in the fit, namely of $\Upsilon$ (10860), $\Upsilon$ (11020), and $\Upsilon$(1D), are predicted with about the same accuracy with which the fitted masses  are reproduced. The remaining differences between the model predictions and the experimental meson masses are of the order of the typical mass splitting between the different spin states, so one cannot expect to do much better with a model without spin dependence. For completeness,  Fig.~\ref{fig:bottomonium} includes also several excited states with $\ell=1$ and $\ell=2$ predicted by this model which have not yet been measured.

We emphasize that the goal of this exercise was not to produce a precision fit of the bottomonium spectrum, but rather to see if our numerical apparatus is reliable and fast enough to make a least-squares fit (and later a least $\chi^2$ fit) of the model parameters possible. Our results are very promising in this respect and point to the possibility of a more extended combined fit of heavy quarkonia, taking the spin dependence into account as well.

\section{Summary and conclusions}
\label{sec:summary}

In this work, we have studied a method to facilitate the use of a linear confining potential in momentum space. It generates a Cauchy principal value singularity, and---in partial waves other than the S-wave---an additional logarithmic singularity, in the kernel of the Schr\"odinger equation. Whereas the logarithmic singularity can be removed by means of a well known subtraction method,  the remaining principal value singularity makes its numerical solution cumbersome.  We showed that this singularity can also be eliminated by another subtraction, thus leaving the resulting equation free of singularities. 

We then demonstrated that in this form the equation is very well suited for a numerical solution. We solved the momentum space Schr\"odinger equation with a linear potential numerically by means of an expansion in a basis of cubic B-splines, and we found that the solutions converge very quickly with increasing number of basis functions. For the S-wave, the exact energies and eigenstate wave functions can be calculated analytically in coordinate space. Our momentum space solutions are in excellent agreement with the exact results. This is true not only for the energy eigenvalues, but also for the wave functions that can be compared after applying a Fourier transform to the coordinate space wave functions.

For higher partial waves no exact solutions are available to test our results. Nevertheless, our singularity-free momentum space equation yields rapid convergence for both radially and orbitally higher excited states. We also solved the case of a screened linear potential and verified that the unscreened limit is reached smoothly and without numerical instabilities.

Finally, we solved the momentum space Schr\"odinger equation with a Cornell-type potential, i.e., a combination of an (unscreened) linear with a Coulomb potential, for bottomonium. We found that our method can be used in practice to perform a least-squares fit of the bottomonium spectrum with this simple model, which reproduces the (spin averaged) experimental masses very well. In doing so we obtained very reasonable values for the potential parameters and for the bottom quark mass. 

The purpose of our work is twofold: First, it is useful to have a practical method at ones disposal to solve the nonrelativistic bound state problem in momentum space for Cornell-type potentials. But our main objective is to extend these calculations to a fully relativistic treatment of quark-antiquark bound states,  in the framework of the Covariant Spectator Theory (CST). Relativistic equations essentially demand a momentum space formulation, and the relativistic generalization of the linear confining potential leads to a covariant CST kernel with the same type of singularity as in the nonrelativistic case. Solving the problem of how to deal with these singularities in the nonrelativistic equation, where the applied techniques can be tested much more easily, therefore paves the way for a practical solution of the relativistic equations. 
In fact, we already performed preliminary studies applying this method to the one-channel CST equation \cite{Gross:1991te,Uzz99} and found it to work as effectively as in the nonrelativistic case, even with retardation. 

Moreover, the CST equation turns into the momentum space Schr\"odinger equation in the nonrelativistic limit. The results of this paper can serve as a benchmark for the relativistic equation, because its solutions should approach the nonrelativistic counterparts with increasing quark masses. It may even turn out sufficient to determine some of the  parameters of the relativistic kernel in the nonrelativistic limit, in particular in fits to the heavy quarkonia, similar to what we have done here in a simplified version for some states of bottomonium.

\begin{acknowledgments}
This work received financial support from Funda\c c\~ao para a Ci\^encia e a Tecnologia (FCT) under grant Nos.~PTDC/FIS/113940/2009, CFTP-FCT (PEst-OE/FIS/U/0777/2013) and POCTI/ISFL/2/275.  The research leading to these results has received funding from the European Community's Seventh Framework Programme FP7/2007-2013 under Grant Agreement No.\ 283286.
\end{acknowledgments}

\appendix
\section{Cauchy principal value integral}
\label{Sec:PV}
The kernel of the momentum-space Schr\"odinger equation for the unscreened linear potential contains a singularity. 
In the literature, one can find demonstrations that the corresponding one-dimensional integral in the partial wave projected equation is of  principal value type \cite{Norbury:1992jv,Spence:1987aa,Spence:1993tb}. However, to give meaning also to  three-dimensional equations involving this singular kernel, such as the three-dimensional  Schr\"odinger equation (\ref{eq:SEa}),  we show here that  the integral over its singular integrand is a three-dimensional Cauchy principal value integral, and as such is well defined.

The potential term of the Schr\"{o}dinger equation for the screened linear potential $ V_{L,\epsilon}({\bf q})$ of Eq.~(\ref{eq:mom}) is   
\begin{eqnarray}
\label{eq:intV2}
I _\epsilon ({\bf p})&=&\int d^3k\, V_{L,\epsilon}({\bf k}-{\bf p}) \Psi ({\bf k})\nonumber\\
&=&\int d^3q \,V_{L,\epsilon}({\bf q}) \Psi ({\bf q}+{\bf p})\nonumber\\&=&
 \int d^3q V_{A,\epsilon}(q)\left[ \Psi ({\bf q}+{\bf p})
- \Psi (\bf p)\right] \, .
\end{eqnarray}
In order to obtain the potential term for the unscreened linear potential, the limit $\epsilon\rightarrow0$ cannot be taken immediately in the integrand, since $V_{A,\epsilon=0}(q)\equiv V_{A}(q)$ is not integrable near ${\bf q}=0$. Instead, the limit  $\epsilon\rightarrow0$ has to be taken \textit{after} the integration. We will show that $ \lim_{\epsilon\rightarrow0}I _\epsilon ({\bf p})\equiv I({\bf p})$ corresponds exactly to a Cauchy principal value. 

For this purpose, it is useful to express $I _\epsilon ({\bf p})$ in terms of spherical coordinates: 
 \begin{equation}
 \label{eq:intV}
I _\epsilon ({\bf p})
=
 \int_{0}^\infty dq\, q^2 V_{A,\epsilon}(q)
 \int d\hat{{\bf q}} \left[ \Psi ({\bf q}+{\bf p})
- \Psi ({\bf p})\right] \, ,
\end{equation}
 where $\hat{ {\bf q}}$ is a unit vector in the direction of ${\bf q}$, and $q=|{\bf p}-{\bf k}|$.
 
Next, writing $q\, V_{A,\epsilon}(q)$ as a derivative,
\begin{eqnarray} 
q \,V_{A,\epsilon}(q)=4\pi \sigma  \frac{d}{d q}\left( \frac{1}{q^2+\epsilon^2}\right)\,,
\end{eqnarray}
and performing one integration by parts, we obtain
\begin{multline}
\label{eq:PVint}
I_\epsilon ({\bf p}) = 4\pi \sigma  \frac{q}{q^2+\epsilon^2}\int d\hat{{\bf q}}
 \left[ \Psi ({\bf q}+{\bf p})
- \Psi ({\bf p})\right]\Big\vert_{q=0}^{\infty} 
\\
  -4\pi \sigma \int_0^\infty d q \, \frac{1}{q^2+\epsilon^2}
\frac{d}{d q} q 
 \int d\hat{{\bf q}} \left[\Psi ({\bf q}+{\bf p})
- \Psi ({\bf p})\right]
  \,.
\end{multline}
The upper limit at $q=\infty$ of the ``surface term'' vanishes provided the wave function satisfies the correct boundary conditions. To realize that the lower limit also vanishes, it is useful to expand the first term in the square brackets into a Taylor series around ${\bf q} =0$,
\begin{eqnarray} \label{eq:expans1}
\Psi ({\bf q}+{\bf p})
- \Psi ({\bf p})=
{\bf q}\cdot {\bf f} ({\bf p})+\mathcal O(q^2)
\end{eqnarray}
where
${\bf f} ({\bf p}) \equiv \nabla_{\bf q}\Psi ({\bf q}+{\bf p})\Big\vert_{{\bf q}=0}$.
We obtain
\begin{equation} \label{eq:expans}
 \int d\hat{{\bf q}} \left[\Psi ({\bf q}+{\bf p})
- \Psi ({\bf p})\right] 
= 
q\,\int d\hat{{\bf q}} \left[\hat{{\bf q}}\cdot  {\bf f}({\bf p})+\mathcal O(q)\right]\,,
\end{equation}
which vanishes in the limit $q \rightarrow0$.  

Using the expansion (\ref{eq:expans1}) again in the remaining integral of Eq.~(\ref{eq:PVint}), we can write 
\begin{equation}
\label{eq:pvint2}
I_\epsilon ({\bf p})  =
-8\pi \sigma   \int_0^\infty d q \frac{q}{q^2+\epsilon^2}
\int  d\hat{{\bf q}}
\left[\hat{{\bf q}}\cdot  {\bf f}({\bf p})+\mathcal O(q)\right]\,.
\end{equation} 

Since  $\int  d\hat{{\bf q}}\, \hat{{\bf q}}\cdot  {\bf f}({\bf p}) =0$, and
the integrand involving the higher-order terms $\mathcal O(q)$ is regular when $\epsilon\rightarrow0$, we conclude that the limit $I ({\bf p})= \lim_{\epsilon\rightarrow0} I _\epsilon ({\bf p})$ exists. 

To show that $I ({\bf p})$ is a Cauchy principal value integral in three dimensions,
first the identity
\begin{equation}
\int_0^\infty d q\frac{q}{q^2+\epsilon^2} 
 = 
 \int_\epsilon^\infty \frac{ d q}{q}
 \label{eq:INTidentity}
 \end{equation} 
is applied to rewrite the integration over $q$ in Eq.~(\ref{eq:pvint2}). (A simple way to see that the identity holds is that the changes of variables $q=\epsilon \sinh y$ on the lhs and $q=\epsilon \cosh y$ on the rhs of Eq.~(\ref{eq:INTidentity}) lead to the same expression). 

Then we apply again (\ref{eq:expans1}), and use $d q \, d\hat{{\bf q}}=d^3 q / q^2$  to substitute the unscreened linear potential $V_A$ back into the integrand. This leads to
\begin{eqnarray}\label{eq:PVint3}
 I({\bf p})&=&
-8\pi\sigma \lim_{\epsilon\rightarrow0} \int_\epsilon^\infty d q \int d\hat{{\bf q}} \,\frac{1}{q^2} \left[ \Psi ({\bf q}+{\bf p})
-  \Psi ({\bf p})\right]\,\nonumber\\&=&
 \lim_{\epsilon\rightarrow0} \int_{q\geq \epsilon} d^3 q \,V_A(q)\left[\Psi ({\bf q}+{\bf p})
- \Psi ({\bf p})\right]\,\nonumber\\&=&
\lim_{\epsilon\rightarrow0} \int_{q\geq \epsilon} d^3 q \,{\bf K}({\bf q})\cdot [{\bf f}({\bf p})+\mathcal O(q)]\,, 
 \end{eqnarray}
 where ${\bf K}({\bf q})={\bf q} V_A(q)$. Since ${\bf K}({\bf q})$ is homogeneous of degree $-3$ and $\int d\hat{{\bf q}}  \,{\bf K}({\bf q})=0$, the last expression is precisely the definition of the Cauchy principal value integral in three dimensions (see, for example, Ref.~\cite{Estrada:2002}). We can therefore write
\begin{eqnarray}
I({\bf p}) &=& 
 \lim_{\epsilon\rightarrow0} \int_{|{\bf k}-{\bf p}|\geq \epsilon} d^3 k \,V_L({\bf k}-{\bf p})\Psi ({\bf k})\nonumber\\
&\equiv& 
\mathrm {P}\!\! \int d^3 k \,V_L({\bf k}-{\bf p})\Psi ({\bf k}) \, ,
\end{eqnarray}
which concludes our proof.

\section{Partial wave decomposition}
\label{Sec:PW}
In this section we outline the partial wave decomposition of Eq.\ (\ref{eq:SEa}) which contains a subtraction term not usually present in the Schr\"odinger equation. In particular, we derive Eqs.\ (\ref{eq:2SPW}) and (\ref{eq:2VAPW}).

Switching to a Dirac notation, we expand the wave function $\Psi({\bf p})$ and the potential $V_A({\bf p},{\bf k})$ into spherical harmonics,
\begin{align}
\Psi({\bf p})=\langle {\bf p} | \Psi \rangle  & = \sum_{\ell' m'} \langle \hat{\bf p} | \ell' m' \rangle \langle p \,  \ell' m' | \psi \rangle \nonumber\\
& = \sum_{\ell' m'} Y_{\ell' m'}(\hat{\bf p}) \langle p \,  \ell' m' | \psi \rangle \, ,
\end{align}
\begin{equation}
\langle {\bf p} | V_A | {\bf k}\rangle 
=
\sum_{n m_n}
\langle p \, n m_n | V_A | k \,  n m_n \rangle 
Y^*_{n m_n}(\hat{\bf k}) Y_{n m_n}(\hat{\bf p}) \, ,
\end{equation}
where $\hat{\bf p}$ is a unit vector in the direction of $\bf p$.

Substituting these expansions into (\ref{eq:SEa}) leads to
\begin{widetext}
\begin{multline}
\frac{p^2}{2m_R} 
\sum_{\ell' m'} Y_{\ell' m'}(\hat{\bf p}) \langle p \,  \ell' m' | \psi \rangle
+  
\sum_{\ell_1 m_1} 
 \sum_{n m_n}
\int_0^\infty  \frac{d k k^2}{(2\pi)^3} \int d \hat{\bf k} \,
\langle p \, n m_n | V_A | k \,  n m_n \rangle 
Y^*_{n m_n}(\hat{\bf k}) Y_{n m_n}(\hat{\bf p})
Y_{\ell_1 m_1}(\hat{\bf k}) \langle k \,  \ell_1 m_1 | \psi \rangle \\
 - 
 \sum_{\ell' m'}
 \sum_{n m_n}
\int_0^\infty  \frac{d k k^2}{(2\pi)^3} \int d \hat{\bf k} \,
\langle p \, n m_n | V_A | k \,  n m_n \rangle 
Y^*_{n m_n}(\hat{\bf k}) Y_{n m_n}(\hat{\bf p})
  Y_{\ell' m'}(\hat{\bf p}) \langle p \,  \ell' m' | \psi \rangle
   =
 E \sum_{l' m'} Y_{l' m'}(\hat{\bm p}) \langle p \,  l' m' | \psi \rangle \, .
\end{multline}
In the first integral over $d \hat{\bf k}$ we use the orthogonality relation
for spherical harmonics,
\begin{equation}
\int d\hat{\bf k}\,  Y^*_{n m_n}(\hat{\bm k}) Y_{\ell_1 m_1}(\hat{\bf k}) = \delta_{n \ell_1}\delta_{m_n m_1} \, ,
\end{equation}
and in the second we use $Y_{00}(\hat{\bf k})=1/\sqrt{4\pi}$ to write
\begin{equation}
\int d\hat{\bf k}\,  Y^*_{n m_n}(\hat{\bf k}) =\sqrt{4\pi} \int d\hat{\bf k}\,  Y^*_{n m_n}(\hat{\bf k}) Y_{00}(\hat{\bf k})
=
\sqrt{4\pi} \delta_{n 0}\delta_{m_n 0} \, .
\end{equation}
The sums over $n$ and $m_n$ can be carried out and give
\begin{multline}
\frac{p^2}{2m_R} 
\sum_{\ell' m'} Y_{\ell' m'}(\hat{\bf p}) \langle p \,  \ell' m' | \psi \rangle
+  
\sum_{\ell_1 m_1} 
\int_0^\infty  \frac{d k k^2}{(2\pi)^3}
\langle p \, \ell_1 m_1 | V_A | k \,  \ell_1 m_1 \rangle 
Y_{\ell_1 m_1}(\hat{\bf p})
 \langle k \,  \ell_1 m_1 | \psi \rangle \\
 - 
 \sum_{\ell' m'}
\int_0^\infty  \frac{d k k^2}{(2\pi)^3} 
\langle p \, 0 0 | V_A | k \,  0 0 \rangle 
\sqrt{4\pi}  Y_{0 0}(\hat{\bf p})
  Y_{\ell' m'}(\hat{\bf p}) \langle p \,  \ell' m' | \psi \rangle
   =
 E \sum_{\ell' m'} Y_{\ell' m'}(\hat{\bf p}) \langle p \,  \ell' m' | \psi \rangle \, .
\end{multline}
In the second integrand, we can simplify again $\sqrt{4\pi}  Y_{0 0}(\hat{\bf p})=1$.

The last step to project out partial wave $(\ell m)$ is to multiply the equation by $Y^*_{\ell m}(\hat{\bf p})$ and integrate over $\hat{\bf p}$. Orthogonality then yields
\begin{equation}
\frac{p^2}{2m_R} 
 \langle p \,  \ell m | \psi \rangle
+  
\int_0^\infty  \frac{d k k^2}{(2\pi)^3}
\Bigl[
\langle p \, \ell m | V_A | k \,  \ell m \rangle 
 \langle k \,  \ell m | \psi \rangle - 
\langle p \, 0 0 | V_A | k \,  0 0 \rangle 
 \langle p \,  \ell m | \psi \rangle
 \Bigr]
   =
 E \langle p \,  \ell m | \psi \rangle \, .
 \label{eq:SPW}
\end{equation}
\end{widetext}

\noindent
Next we calculate the partial wave matrix elements of $V_A$,
\begin{multline}
\langle p \, \ell m | V_A | k \,  \ell m \rangle =
2\pi  (-8\pi\sigma)
\int_{-1}^1 dx \frac{P_\ell(x)}{(k^2+p^2-2pk x)^2} \\
= 
2\pi  \frac{(-8\pi\sigma)}{(2pk)^2}\int_{-1}^1 dx \frac{P_\ell(x)}{(\frac{k^2+p^2}{2pk}- x)^2} \, ,
\label{eq:VAPW}
\end{multline}
where $x \equiv \hat{\bf p}\cdot \hat{\bf k}$.
Introducing the abbreviation
\begin{equation}
y=\frac{p^2+k^2}{2pk} \, ,
\end{equation}
and using the Legendre functions of the second kind,
\begin{equation}
Q_\ell(y)= \frac{1}{2} \int_{-1}^1 dx \frac{P_\ell(x)}{y-x} \, ,
\end{equation}
we can write
\begin{equation}
\int_{-1}^1 dx \frac{P_\ell(x)}{(y-x)^2}=
-\frac{d}{d y}  \int_{-1}^1 dx \frac{P_\ell(x)}{y-x}
 = -2 Q'_\ell(y) \, .
\end{equation}
For the case $\ell=0$ we obtain
\begin{equation}
Q_0(y) = \frac{1}{2} \ln \left| \frac{y+1}{y-1} \right| 
= \frac{1}{2} \ln \left( \frac{p+k}{p-k}\right)^2 \, ,
\end{equation}
and
\begin{equation}
Q'_0(y) =\frac{1}{1-y^2}=-\frac{4 p^2 k^2}{\left( p^2-k^2 \right)^2} \, .
\end{equation}
Both $Q_0(y)$ and $Q'_0(y)$ are singular at $k=p$. For higher partial waves we can use the relation
\begin{equation}
Q_\ell(y) = P_\ell(y) Q_0(y) -w_{\ell-1}(y) \, ,
\end{equation}
where
\begin{equation}
w_{\ell-1}(y) = \sum_{m=1}^\ell \frac{1}{m} P_{\ell-m}(y) P_{m-1}(y) \, .
\end{equation}
This shows that---in all partial waves---the only singularities come from $Q_0(y)$ and $Q'_0(y)$.

The partial wave matrix element of $V_A$ can now be expressed as
\begin{align}
& \langle p \, \ell m | V_A | k \, \ell m \rangle 
\nonumber\\
& = 2\pi 
\frac{(-8\pi\sigma)}{(2pk)^2}(-2)
\left[ 
 P_\ell(y) Q'_0(y)
 +
 P'_\ell(y) Q_0(y) 
 -w'_{\ell-1}(y) \right] 
\nonumber \\
& = 2\pi (-8\pi\sigma)
\Biggl[ 
\frac{2 P_\ell(y)}{\left( p^2-k^2 \right)^2}
-
\frac{P'_\ell(y)}{\left( 2p k \right)^2} \ln \left( \frac{p+k}{p-k}\right)^2
\nonumber\\
&\qquad\qquad\qquad +
\frac{2 w'_{\ell-1}(y)}{\left( 2p k \right)^2}
\Biggr] .
\label{eq:VAPWfinal}
\end{align}
The matrix elements are independent of $m$ due to rotational symmetry, so (\ref{eq:VAPW}) is the same as (\ref{eq:2VAPW}).
After substituting (\ref{eq:VAPWfinal}) into the partial-wave Schr\" odinger equation (\ref{eq:SPW}) we can sum over $m$ and divide by $2\ell+1$, which finally yields Eq.~(\ref{eq:3SPW}).

\bibliographystyle{h-physrev3}
\bibliography{PapersDB-v1.4.2}

\end{document}